\documentclass{article}

\usepackage{arxiv}

\usepackage[utf8]{inputenc} 
\usepackage[T1]{fontenc}    
\usepackage{hyperref}       
\usepackage{url}            
\usepackage{booktabs}       
\usepackage{amsfonts}       
\usepackage{nicefrac}       
\usepackage{microtype}      
\usepackage{multirow}
\usepackage{lipsum}
\usepackage{graphicx}
\usepackage{amsmath}
\usepackage{nccmath}
\usepackage[super]{nth}
\usepackage{float}
\graphicspath{ {./images/} }

\title{Predicting Road Flooding Risk with Machine Learning Approaches Using Crowdsourced Reports and Fine-grained Traffic Data}

\author{
 Faxi Yuan \\
  Urban Resilience.AI Lab\\
  Zachry Department of Civil and \\Environmental Engineering\\
  Texas A\&M University\\
  College Station, TX 77843 \\
  \texttt{faxi.yuan@tamu.edu} \\
  \And
 William Mobley \\
  Institute for a Disaster Resilient Texas\\
  Department of Marine Sciences\\
  Texas A\&M University at Galveston\\
  Galveston, TX 77553 \\
  \texttt{wmobley@tamu.edu} \\
  \And
 Hamed Farahmand \\
  Urban Resilience.AI Lab\\
  Zachry Department of Civil and \\Environmental Engineering\\
  Texas A\&M University\\
  College Station, TX 77843 \\
  \texttt{hamedfarahmand@tamu.edu} \\
  \And 
 Yuanchang Xu \\
  Urban Resilience.AI Lab\\
  Department of Computer Science \\and Engineering\\
  Texas A\&M University\\
  College Station, TX 77843 \\
  \texttt{yuanchangxu@tamu.edu} \\
  \And
 Russell Blessing \\
  Institute for a Disaster Resilient Texas\\
  Department of Marine Sciences\\
  Texas A\&M University at Galveston\\
  Galveston, TX 77553 \\
  \texttt{rblessing@tamug.edu} \\
  \And
 Shangjia Dong \\
  Department of Civil and \\Environmental Engineering\\
  University of Delaware\\
  Newark, DE 19716 \\
  \texttt{sjdong@udel.edu} \\
  \And
 Ali Mostafavi \\
  Urban Resilience.AI Lab\\
  Zachry Department of Civil and \\Environmental Engineering\\
  Texas A\&M University\\
  College Station, TX 77843 \\
  \texttt{amostafavi@civil.tamu.edu} \\
  \And
 Samuel D. Brody \\
  Institute for a Disaster Resilient Texas\\
  Department of Marine Sciences\\
  Texas A\&M University at Galveston\\
  Galveston, TX 77553 \\
  \texttt{brodys@tamug.edu} \\
}

\begin{document}
\maketitle
\begin{abstract}
The objective of this study is to predict road flooding risks based on topographic, hydrologic, and temporal precipitation features using machine learning models. Predictive flood monitoring of road network flooding status plays an essential role in community hazard mitigation, preparedness, and response activities. Existing studies related to the estimation of road inundations either lack observed road inundation data for model validations or focus mainly on road inundation exposure assessment based on flood maps. This study addresses this limitation by using crowdsourced and fine-grained traffic data as an indicator of road inundation, and topographic, hydrologic, and temporal precipitation features as predictor variables. Two tree-based machine learning models (random forest and AdaBoost) were then tested and trained for predicting road inundations in the contexts of 2017 Hurricane Harvey and 2019 Tropical Storm Imelda in Harris County, Texas. The findings from Hurricane Harvey indicate that precipitation is the most important feature for predicting road inundation susceptibility, and that topographic features are more essential than hydrologic features for predicting road inundations in both storm cases. The random forest and AdaBoost models had relatively high AUC scores (0.860 and 0.810 for Harvey respectively and 0.790 and 0.720 for Imelda respectively) with the random forest model performing better in both cases. The random forest model showed stable performance for Harvey, while varying significantly for Imelda. This study advances the emerging field of smart flood resilience in terms of predictive flood risk mapping at the road level. For example, such models could help impacted communities and emergency management agencies develop better preparedness and response strategies with improved situational awareness of road inundation likelihood as an extreme weather event unfolds.
\end{abstract}

\keywords{smart resilience \and road flood risk \and machine learning \and big data \and urban flood}

\section*{Introduction}

Road networks play a critical role in the transportation of goods, access to food and healthcare, and economic activities (Pregnolato et al. 2017). Road inundations during major flood and storm events reduce the access of impacted communities to essential facilities such as hospitals (Dong et al., 2020a) and grocery stores (Podesta et al., 2021), and present challenges for emergency management agencies to prepare, design and implement response strategies (Yuan et al., 2021a). In addition, drivers may unsuccessfully try to navigate flooded urban roads, resulting in loss of life when rescue efforts fail (Drobot et al., 2007). Therefore, the ability to predict road inundations in advance of hurricanes and floods provides emergency management agencies enhanced regional situational awareness by identifying the highly likely loss of access to critical facilities, such as hospitals and shelters (Dong et al., 2020a). This study aims to create and test machine learning models for predicting road inundation probabilities based on their topographic, hydrologic, and temporal precipitation features by using crowdsourced Waze reports and fine-grained traffic data as indications of road inundations. The rest of this section will discuss the extant literature related to flood risk modeling and their limitations in terms of predicting road inundation probabilities to establish the point of departure for this study.

\subsection*{Hydraulic and hydrologic models for urban flood inundations}

Various studies have focused on urban flood inundation with hydraulic and hydrologic (H\&H) models (Chen et al., 2018; Jamali et al., 2018; Cook and Merwade, 2009), such as 1-D modeling with HEC-RAS (Chaudhry et al., 2018), and 2-D modeling with an urban inundation model (Chen et al., 2007) and LISFLOOD‐FP (Bates and de Roo, 2000). However, H\&H models, such as an urban inundation model and LISFLOOD‐FP need to solve the full shallow water equations (SWEs) and further require a considerable amount of computation resources (Jamali et al., 2019). Given the complexity and high computational demands of these models, recent studies have attempted to build models that do not resolve SWEs. Jamali et al., (2019) categorized these latter H\&H models into two categories based on their complexity: models based on cellular automata (CA) and models based on topographic depressions. CA-based models divide flood domains into a set of regular grid cells and require small time steps for flood inundation simulations but are also computationally intensive (Liu et al., 2015). Models based on topographic depressions are referred to as rapid flood models in Jamali et al., (2019), and depend mainly on topographic features and the continuity equation for urban flood inundation simulation but lack temporal features such as precipitation (Nguyen and Bae, 2020). Studies integrating H\&H models with machine learning approaches (Hou et al., 2021) use the flood depth outputs from hydrodynamic models as training datasets and rainfall data as the primary predictor (Hou et al., 2021), however limited observed urban flood inundation data makes validation difficult (Smith et al., 2012). For instance, Lyu et al., (2019) used limited public reports of flood incidents from websites such as Google and Baidu and from literature (Huang et al., 2017; Yin et al., 2016b) to validate their simulated urban flood inundations in Shanghai. Another limitation for H\&H models is that outputs refer mainly to the general pattern of flood inundations over large metropolitan areas  (Lyu et al., 2019; Yu et al., 2016), while struggling to accurately predict small-scale flooding such as road inundations.

Despite these limitations, H\&H models have been extensively used to estimate road inundations (e.g., Versini, 2012). For example, Coles et al., (2017) employed the hydrodynamic flood inundation model (FloodMap) to simulate two pluvial flood events in York, UK and then, identified the regions with restricted accessibility for emergency responders. Using a road network analysis, their study evaluated emergency service accessibility. Yin et al., (2016a) extended this approach by integrating a hydrodynamic model (FloodMap HydroInundation 2D) and flood depth-dependent measures to assess the road inundations in a pluvial flash flood event in Shanghai, China. Their hydrodynamic model was based on rainfall scenarios from the intensity–duration–frequency relationships of a Shanghai rainstorm and the Chicago Design Storm. But again, their simulated flood inundations cannot be validated with observed flood inundation data. A prior study by Versini (2012) attempted to overcome this limitation by using historical road inundation data to define four road inundation risk levels (high, medium, low, and safe). These risk levels were then evaluated against simulated discharges from a hydrological model and used to establish a real-time flood warning system. However, the models in these studies provide insight only about road inundation exposure based on historical road inundations; they do not determine a continuous road-level inundation probability. This limitation is due partly to the dearth of road-level inundation data to verify the results road exposure insights obtained from H\&H models (e.g., Hou et al., 2021; Lyu et al., 2019; Smith 2012). Meanwhile, such a limitation is also due partly to the lack of integration of topographic features of roads (such as elevation and proximity to stream) with the water depth and velocity estimates from H\&H models (Versini 2012). This limitation could be potentially addressed with the use of crowdsourced and fine-grained traffic data that provide reliable indications regarding the inundation status of road sections during storm events and with the development of topographic characterization of roads.

\subsection*{Point of departure}

In this study, we aim to predict the inundation likelihood of road sections based on their topographic, hydrologic, and temporal precipitation features. Existing studies using H\&H models have identified many of the essential features for flood inundations. For instance, Karaoui et al., (2018) highlighted the importance of land use and land cover variations for predicting flood intensity. Whereas Nguyen and Bae (2020) and Yin et al. (2016a) discussed the importance of precipitation for estimating urban flood inundation. Finally, Lee et al. (2017) indicated the critical role of topographic features, such as distance to rivers, for predicting flood inundation. 

This study proposes to use machine learning models to predict road inundations. Similarly to the methods used by Mobley et al., (2021), Nguyen and Bae (2020), Karaoui et al., (2018), and Lee et al., (2017), we employ three categories of input features for random forest and AdaBoost models: (i) topographic features (e.g., proximity to streams and coastlines); (ii) hydrologic features (e.g., land surface roughness); and (iii) temporal precipitation features. Referring to Lyu et al., (2019) and Yu et al., (2016), we use crowdsourced and traffic sensor data to detect road inundation status as a dependent variable for these machine learning models. Using 2017 Hurricane Harvey and 2019 Tropical Storm Imelda in Harris County, Texas, as case studies, we train and test these two models for predicting road inundations.

\section*{Methods and Materials}

Ten independent variables within three feature categories—topographic, hydrologic, and precipitation features—were created using secondary data sources, and the dependent variable, road inundation status, was developed using Waze reports and IRINX traffic data. Due to limited data availability, we used IRINX traffic data for the Harvey and Waze reports for Imelda. While this constraint could produce some limitations in comparing results, it provides an opportunity to compare model results obtained based on sensor data versus crowdsourced data. In particular, we compared the model prediction stability for each case. Two tree-based methods, including random forest and AdaBoost models, were employed for the prediction of road inundation. The framework is illustrated in Figure 1. To validate the implementation of the framework, we introduced two case studies of flood events in Harris County: Hurricane Harvey in 2017 and Tropical Storm Imelda in 2019. Using these two cases, we trained and tested our machine learning models for the prediction of road inundation likelihood.

\begin{figure}[ht]
\centering
\includegraphics[width=0.85\linewidth]{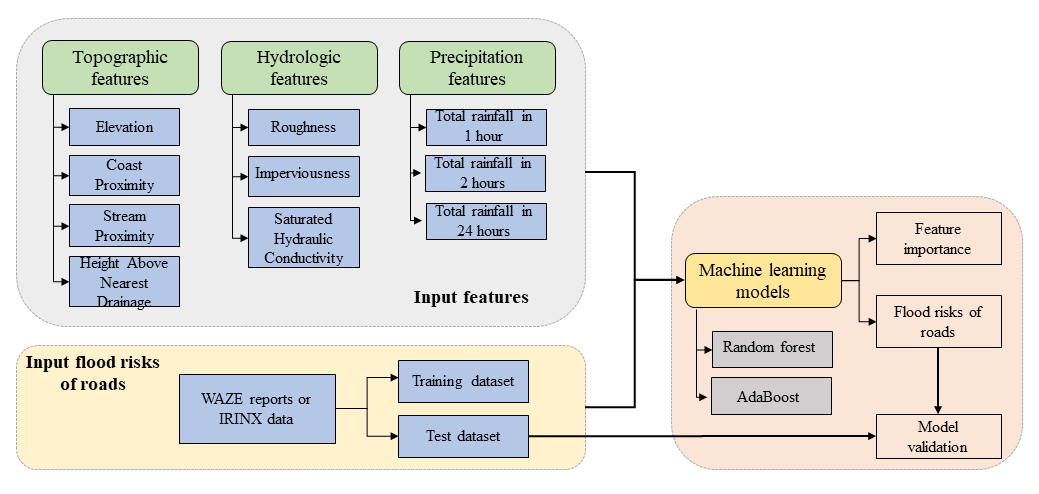}
\caption{Road flood risk prediction framework of machine learning models.}
\end{figure}

\subsection*{Case study region}
Harris County, home to Houston, the fourth largest city in the United States, has experienced rapid population growth over the past decades (Qian 2010). Harris County is also among the most flood-prone counties in the United States due to its coastal location, burgeoning urban development, and the lack of flood control infrastructure development in parallel with the development and population growth (Dong et al., 2020b). Accordingly, the county has experienced several severe flood events, including the Memorial Day Flood in 2015, the Tax Day Flood in 2016, and Hurricane Harvey in 2017. Each event caused extensive flooding with losses ranging from hundreds of millions to hundreds of billion dollars (Dong et al., 2020c). Therefore, Harris County was selected as the testbed for this study. Hurricane Harvey, which made landfall in 2017, caused one of the most devastating floods experienced by Harris County. A Category 4 storm, Hurricane Harvey landed on the Texas Gulf Coast on August 25, 2017, and dissipated inland on August 30, 2017. Hurricane Harvey had extensive economic and social consequences (NOAA, 2017). Similarly, Tropical Storm Imelda caused one of the most devastating floods in decades experienced in Southeast Texas. After landfall throughout the morning September 19, 2019, widespread flooding occurred in Harris County, with over 40 inches (102 centimeters) of rainfall recorded, many of the local rivers and bayous overflowed and inundated a vast area in the county. During Tropical Storm Imelda, more than 1,000 people requested rescue, and caused numerous injuries, deaths, and billions of dollars of losses in Southeast Texas.

\subsection*{Data and feature descriptions}
\subsubsection*{\textit{Input features}}

Figure 1 shows the three categories of features used for predicting road flood risk: topographic, hydrologic, and precipitation features. Topographic features are elevation, coastal and stream proximity, and height above nearest drainage. Hydrologic features include roughness, imperviousness of surface, and the saturated hydraulic conductivity. Both topographic and hydrologic features are generally used in the H\&H models for flood inundation simulations (Nguyen and Bae, 2020; Karaoui et al., 2018; Lee et al., 2017). Temporal precipitation comprises total rainfall in 1 hour, 2 hours, and 24 hours, during a time period when the roads were detected as flooded. The following introduces how these features can be collected and computed.

\textbf{Elevation}: The elevation variable is frequently used to model flood hazard (Lee et al., 2017; Tehrany et al., 2019; Dodangeh et al., 2020; Janizadeh et al., 2019; Rahmati \& Pourghasemi, 2017; Bui et al., 2019; Hosseini et al., 2020; Mobley et al., 2019; Darabi et al., 2019). Higher elevation areas drain to lower areas making these low-lying areas more susceptible to flooding and ponding during rainfall events. Elevation was collected from the USGS 3DEP (United States Geological Survey 3D Elevation) program at 10-meter resolution.

\textbf{Coastal and stream proximity}: Proximity to streams and the coastline are significant predictors of flood damage (Brody et al., 2015). Streams will often overbank during large precipitation events and areas closer to the coast are often prone to storm surge. Both distance to streams and coast were calculated based on the National Hydrography Dataset (NHD) stream and coastline features. 

\textbf{Height above nearest drainage}: While the elevation variable is used to identify high and low areas in relation to sea level, height above nearest drainage (HAND) defines the height of a location above the nearest stream (Garousi-Nejad et al., 2019). Areas of higher HAND values should be less likely to flood due higher precipitation reach those location. HAND is commonly used for calculating flood depths, insured losses from floods (Rodda, 2005), soil water potential (Nobre et al., 2011), groundwater potential (Rahmati and Pourghasemi 2017), and flood potential (Nobre et al., 2011). The HAND dataset was downloaded from the University of Texas’ National Flood Interoperability Experiment continental flood inundation mapping system (Liu et al., 2016) at 10-meter resolution. 

\textbf{Roughness}: Roughness impacts stormwater flow speed as well as the peak flow within channels (Acrement and Schneider, 1984). Overland water flow is often approximated using Manning’s roughness coefficient (Anderson et al., 2006; Thomas and Nisbet, 2007). This coefficient, called Manning’s n, has become a critical input in many hydrological models and has also been shown be a good predictor of event-based flood susceptibility (Mobley et al., 2019). For this study, roughness values were assigned to each National Land Cover Database (NLCD) land cover class using the values suggested by Kalyanapu et al., (2009) and was averaged across the contributing upstream area for each raster cell.

\textbf{Imperviousness}: Imperviousness is another strong indicator of water infiltration. Increasing areas impervious surfaces due of urbanization reduces infiltration and causes increased surface runoff and larger peak discharges (Anderson, 1970; Hall 1984; Arnold and Gibbons, 1996; White and Greer, 2006). Imperviousness is another strong predictor of flooding in the Houston region due to the landscape’s sprawling nature (Brody et al., 2015; Gori et al., 2019; Sebastian et al., 2019; Lee and Gharaibeh, 2020). For this study, impervious was measured using the percent developed impervious surface raster from the 2016 NLCD. Values range from 0\% to 100\% and represent the proportion of urban impervious surface within each 30-m cell.

\textbf{Saturated hydraulic conductivity}: Soil infiltration determines the speed and amount of precipitation absorbed by the soil profile. When stormwater cannot easily percolate through the soil profile, excessive runoff results, particularly in urbanized and downstream areas. Both measures of soil infiltration lithology and saturated hydraulic conductivity (Ksat), are strong predictors of flood susceptibility (Brody et al., 2015; Janizadeh et al., 2019; Bui et al., 2019; Hosseini et al., 2020; Mobley et al., 2019). For this study, Ksat values were assigned to soil classes obtained from the Natural Resources Conservation Service’s Soil Service Geographic Database using the values presented in Rawls et al. (1983), and then averaged across the upstream contributing area for each cell.

\textbf{Precipitation features}: Rainfall totals (in millimeters) are heterogeneous both spatially and temporally during a rain event. High-intensity precipitation over the short term may produce flash-flooding along streets, while longer continuous rain may cause longer-duration inundation. This study calculates total rainfall at one-hour intervals for periods of 1 hour, 2 hours, and 24 hours prior to detection of a flooded road within the dataset using the National Weather Service Gauge Corrected Quantitative Precipitation Estimate dataset. (Data was initially located at the National Oceanic and Atmospheric Administration and is now archived in the Iowa State University database.)

\subsubsection*{\textit{Road Flood Status}}
\textbf{Waze reports data in Tropical Storm Imelda}: Tropical Storm Imelda, the fifth-wettest tropical cyclone on record in the United States, made landfall in Harris County on September 19, 2019. It caused devastating and record-breaking floods and numerous injuries and deaths in Southeast Texas. Mobile navigation applications such as Waze are equipped with real-time information-reporting panels that enable collecting incident reports ranging from road closures and traffic congestion to weather-related hazards, such as flooding, that impact the functionality of roads. This capability enables acquisition and sharing of aggregated user-reported incident data to enhance situational awareness. For assessment of predictive road flooding during Tropical Storm Imelda, we used Waze time-stamped and location-specific flood incident reports. We collected the flooding reports generated by Waze users during the course of Imelda in Harris County. Waze is a GPS navigation app that allows drivers to report road conditions, including flooding. Each report is characterized by the location and a time stamp. A total 41,501 reports indicating weather hazard or road closure were registered in the study area over the 5-day period that Imelda impacted Harris County. These reports indicate flooding alerts in 4,980 road segments in the study area (red nodes in Figure 2a). For the training dataset, we used the random sample method for the random selection of equivalent-size non-flooded roads from county roads data. Our training and test dataset for Imelda includes 9,960 road segments.

\textbf{IRINX traffic data in Hurricane Harvey}: Hurricane Harvey approached Harris County on August 25 and caused severe floods from August 26, 2017 to September 4, 2017. Our study period extends one week before August 27 and one week after September 4. From August 20, 2017 to September 11, 2017, we collected the traffic condition data for 19,712 road segments in Harris County from the INRIX, a private company that provides location-based data and analytics. INRIX collects timely traffic data, such as average traffic speed, matched with road locations through sensors and from connected vehicles. The INRIX traffic data includes the average traffic speed on individual road segments at 5-minute intervals and the segments’ corresponding historical average traffic speed. The INRIX traffic data includes all available road segments the road hierarchy from interstate highways to neighborhood streets in Harris County. Road segment are identified by road name, geographic locations specified by head and end coordinates (latitudes and longitudes), and length. Recent studies have used the IRINX traffic data for flood risk predictions. Road segments with null value as average traffic speed were assumed to be flooded in Hurricane Harvey (Yuan et al., 2021b, 2021c; Fan et al., 2020). As we also considered the temporal features for predicting road flooding risks, we recorded as flooded the period of time during which the road segments had a null value for average traffic speed. We collected data on 1,063 flooded roads during Hurricane Harvey. The balance of road segments, 18,649, roads were presumed to be non-flooded roads. The distribution of flooded and non-flooded roads is illustrated in Figure 2b (red nodes for roads with flood risks and green nodes for non-flooded roads). To create a balanced dataset for implementing the random forest model, we utilized the random sample function to select the equivalent-size dataset of non-flooded roads. In total, we built the dataset with 2,126 roads from the IRINX traffic data. 

\begin{figure}[ht]
\centering
\includegraphics[width=1.0\linewidth]{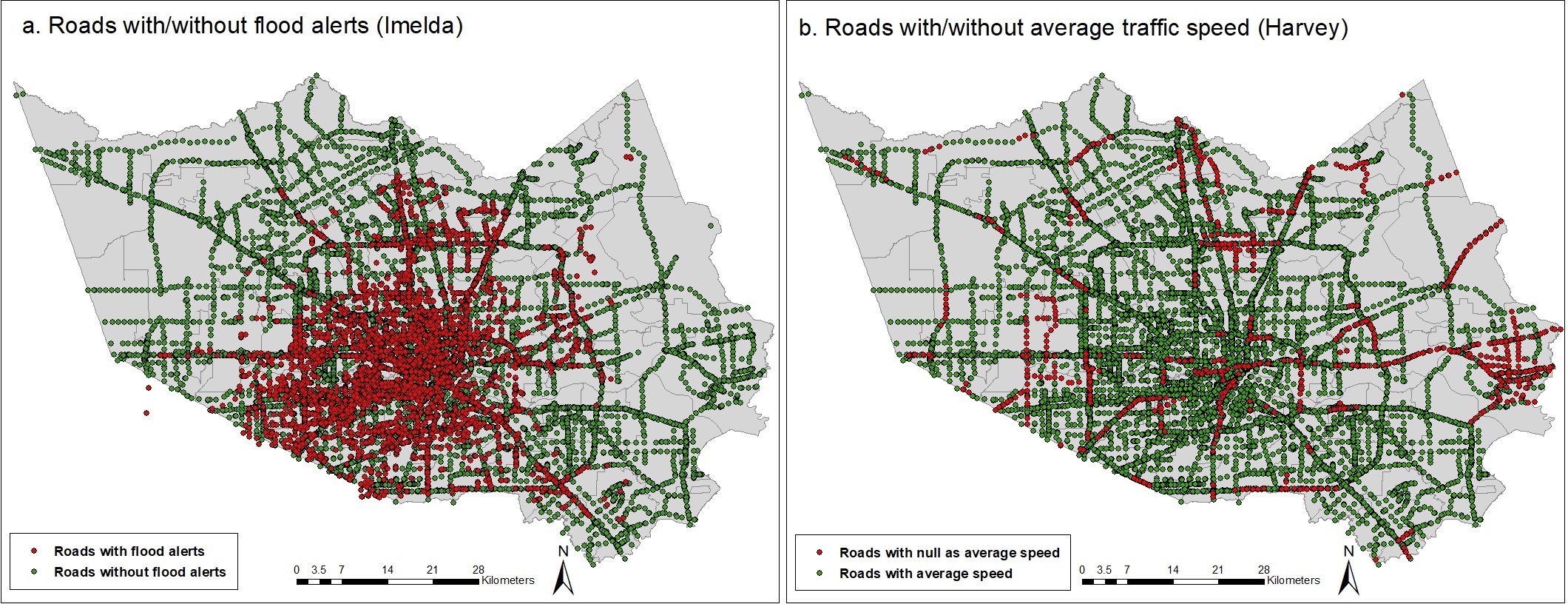}
\caption{Geographic distributions of flooded and non-flooded roads from Waze reports for Tropical Storm Imelda (2a) and from IRINX for Hurricane Harvey (2b). Each node represents a location of roads (green: roads without flood warns or showing average traffic speed; red: roads with flood warnings or null traffic speed).}
\end{figure}

\subsection*{Machine learning models}
Based on the concept of ensemble learning, two common techniques, bagging and boosting, were proposed for the tree-based models (Sutton 2005). Decision tree models that use bagging and boosting techniques consist of multiple decision trees for categorization, which often reduce the errors and variances inherent in a single decision tree (Oza and Russell 2011). The bagging technique divides the initial training dataset into several subsets and then choses them randomly with replacement to train their corresponding decision trees. As a result, the bagging technique produces an ensemble of different tree models. Extended from the bagging technique, random forest was developed by introducing the random selection of features within the training dataset (Prasad et al., 2006). In this study, we implemented the random forest model to predict the road flooding risk based upon their topographic, hydrologic, and precipitation features.

To properly use the random forest, we tuned two critical parameters to reduce error rates, the number of trees and the tree depth. The number of trees define the forest size. Increasing the forest size can reduce errors and involve more features for decisions (Liaw and Wiener 2002); however, such increase requires a greater computational demand. Tree depth refers to the longest path between the root node and the leaf node. The greater the tree depth, the more splits are expected, which captures more information from the feature data; however, a tree that is too deep could result in overfitting. Referring to Mobley et al. (2021), we initially set the number of trees to 200 and tree depth to 90 in our model.

To enable variable selection for enhancing the generalizability of models for predicting road inundation, we employed the feature importance function. Specifically, we used the aggregated decrease in Gini impurity to evaluate feature importance. A greater aggregated decrease in the Gini impurity signifies more important role of the feature. The calculations of feature importance are available from Li et al. (2021).

AdaBoost is another decision-tree-based model that implements a boosting technique for predicting road flooding risk (Schapire 2013). In contrast to the subset replacement method of random forest model, boosting uses the same dataset to build the decision trees for all iterations and revises the weights of inputs in each iteration. The boosting technique analyzes the data of a simple decision tree for errors. Consecutive trees increase the weight of an input misclassified by the previous tree and are more likely to classify it correctly. As a result, the boosting technique output is an ensemble of different tree models. As the first successful implementation of boosting technique for binary classification, AdaBoost has demonstrated strong predictive power for flood risk (Liu et al., 2017; Coltin et al., 2016). In this research, we compared the performances of AdaBoost with that of random forest model for predicting road flood risks with our defined features.

In addition, this research implemented the 10-iteration 5-fold cross-validation process to evaluate the performances of random forest and AdaBoost models for predicting road flooding risk in both Hurricane Harvey and Tropical Storm Imelda. Specifically, we defined the sizes of training and test datasets as 80\% (for training and validation) and 20\% of our initial datasets for both cases. To maintain the high-level randomness for each-fold split of training and test datasets, we employed the train\_test\_split function from the scikit-learn library. To evaluate the performances of random forest and AdaBoost models, we used average accuracy, and average area under the curve (AUC) of the receiver operating characteristic (ROC) in the 10-iteration 5-fold cross-validation process. In this study, we used the flooded roads as positive class for the probability predictions. Accuracy reflects the percentage of correctly predicted roads with flood risks and those with non-flood risks (Eq. (1)). Sensitivity denotes the percentage of correctly predicted roads with flood risks (Eq. (2)). The AUC of the ROC reveals the estimates of the probability that the models will correctly predict flooded roads as roads with flood risks. With the prediction results, the ROC curve could be defined by the relationship between true positive and false positive rates (Eq. (2) through (4)).  We also want to note that all the random forest and AdaBoost models were trained and tested with the scikit-learn library in Python version 3.7.

\begin{equation}
\label{eq:1}
Accuracy= \frac{True \ positive+True \ negative}{True \ positive +True \ negative +False \ positive+False \ negative}
\end{equation}

\begin{equation}
\label{eq:2}
Sensitivity= True \ positive \ rate= \frac{True \ positive}{True \ positive +False \ positive}
\end{equation}

\begin{equation}
\label{eq:3}
Specificity= \frac{True \ negative}{True \ negative +False \ positive}
\end{equation}

\begin{equation}
\label{eq:4}
False \ positive \ rate= 1 -Specificity = \frac{False \ positive}{True \ negative +False \ positive}
\end{equation}

where \textit{true positive} denotes the situation where the models correctly predicted the road flooding risk, while \textit{true negative} is for the result of correct prediction of non-flood risks of roads; \textit{false positive} refers to the outcome where the models incorrectly predicted the road flooding risk, while \textit{false negative} is for the result that the models incorrectly predicted the non-flooded status of roads. 

\section*{Results}
\subsection*{Feature importance for random forest}

This section illustrates the rank of feature importance for random forest model for Tropical Storm Imelda (Figure 3a) and Hurricane Harvey (Figure 3b). A significant difference between these ranks is the rank of precipitation features. The precipitation features are a stronger indicator of road flood risk during Hurricane Harvey than that in Tropical Storm Imelda. According to the National Hurricane Center reports, the highest rainfall totals were 44.49 inches (1122.43 mm) Tropical Storm Imelda (Latto and Berg 2020) and 60.58 inches (1538.73 mm) for Hurricane Harvey (Blake and Zelinsky 2018). These reports indicate that rainfall volume brought by Hurricane Harvey is much larger than that of Tropical Storm Imelda. Accordingly, rainfall could have resulted in more severe flood risks in Hurricane Harvey. As a result, precipitation features have higher ranks of importance in Hurricane Harvey than in Tropical Storm Imelda.

\begin{figure}[ht]
\centering
\includegraphics[width=\linewidth]{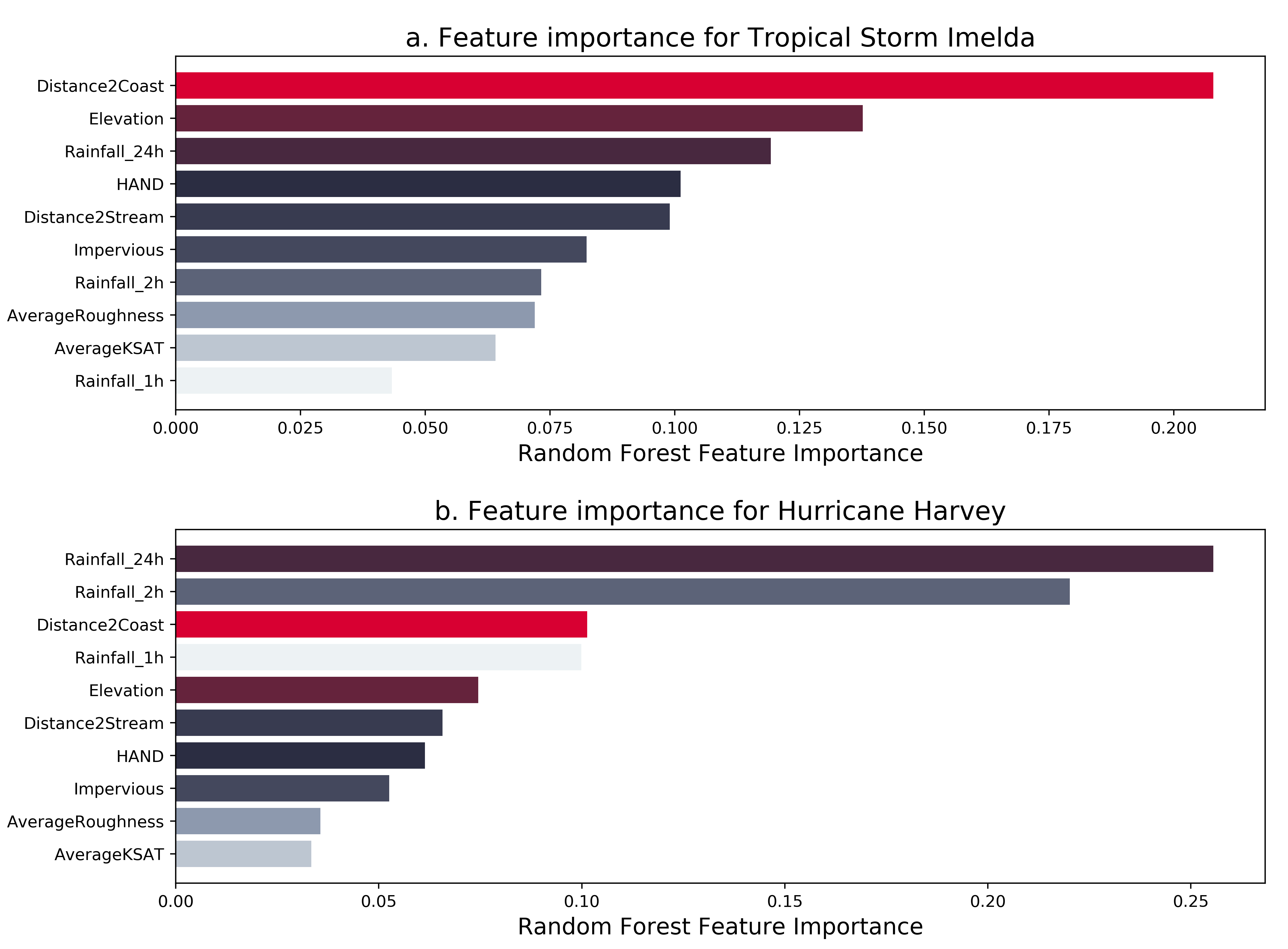}
\caption{Feature importance results for Tropical Storm Imelda (3a) and Hurricane Harvey (3b) with random forest. A larger value of feature importance indicates a more important role of that feature.}
\end{figure}

In addition to precipitation features, ranks of topographic and hydrologic features remain almost constant for both Tropical Storm Imelda and Hurricane Harvey, excluding the ranks of stream proximity (i.e., Distance2Stream) and height above nearest drainage (HAND). The general trend is that topographic features are stronger predictors of road inundation than hydrologic features for both events using the random forest model. Specifically, factors influencing the infiltration of stormwater into the ground, such as saturated hydraulic conductivity of soil (AverageKSAT), impervious surface (impervious), and roughness (AverageRoughness), were poor predictors of road inundation in both Tropical Storm Imelda and Hurricane Harvey, whereas factors influencing where water tends accumulate such as Elevation, HAND, Distance2Coast, Distance2Stream, were all strong predictors of road inundation. This finding is consistent with existing studies, as elevation is one of the frequently used factors for flood hazard simulations (Mobley et al., 2019; Darabi et al., 2019), and proximity to a coast was found to be an strong indicator of flood damage (Brody et al., 2015). 

\subsection*{Model performances of random forest and AdaBoost}
With the 10-iteration 5-fold cross-validations, we computed the average of accuracy and sensitivity for the model performance (Table 1). We also recorded the variations of each evaluation matrix in the process (accuracy and AUC ranges in Table 1). The random forest model had better performance for predicting road flooding risks for both Tropical Storm Imelda and Hurricane Harvey than AdaBoost model in terms of accuracy (Table 1). The random forest model demonstrated higher accuracy (0.900 versus 0.764) for predicting road inundation in Hurricane Harvey than that for Tropical Storm Imelda, which is the same as observed in the AdaBoost model.

\begin{table}[]
\centering
\caption{Results of evaluation matrix for the model performances}
\begin{tabular}{lllll}
\hline
\multirow{2}{*}{Evaluation matrix} & \multicolumn{2}{l}{Random   forest}      & \multicolumn{2}{l}{AdaBoost}             \\ \cline{2-5} 
                                   & Tropical Storm Imelda & Hurricane Harvey & Tropical Storm Imelda & Hurricane Harvey \\ \hline
Accuracy                           & 0.764                 & 0.900            & 0.689                 & 0.857            \\
Accuracy (range)                   & 0.755   ± 0.015       & 0.895   ± 0.015  & 0.690   ± 0.020       & 0.855   ± 0.025  \\
AUC                                & 0.790                 & 0.860            & 0.720                 & 0.810            \\
AUC (range)                        & 0.   790 ± 0.040      & 0.860   ± 0.100  & 0.720   ± 0.060       & 0.810   ± 0.140  \\
Prediction ability                 & Acceptable            & Excellent        & Acceptable            & Excellent        \\ \hline
\end{tabular}
\end{table}

Using the average of the calculated results of true positive and false positive rates from the 10-iteration 5-fold cross-validation process, we created ROC curves from random forest and AdaBoost models for both events (Figure 4). The random forest and Adaboost average AUC was 0. 790 ± 0.040 and 0.720 ± 0.060 respectively for Tropical Storm Imelda, and 0.860 ± 0.100 and 0.810 ± 0.140 respectively for Hurricane Harvey (Table 1). In other words, the chance of the random forest model correctly predicting road with a high probability of being inundated in Hurricane Harvey is 86.0\%, while that of AdaBoost model is 81.0\%. Meanwhile, the random forest model had a probability of 79.0\% to accurately predict road flooding risk in Tropical Storm Imelda; AdaBoost model had only 72.0\%. Referring to Hosmer and Lemeshow (2000), the ability of a model to predict roads with and without flood risk can be assigned to one of four level by AUC: 1) AUC = 0.5: no prediction ability; 2) 0.7 $\leq$ AUC $\leq$ 0.8: acceptable; 3) 0.8 $\leq$ AUC $\leq$ 0.9: excellent; and 4) AUC $>$ 0.9: outstanding. Accordingly, the prediction ability of both models for both cases is presented in Table 1.

\begin{figure}[ht]
\centering
\includegraphics[width=\linewidth]{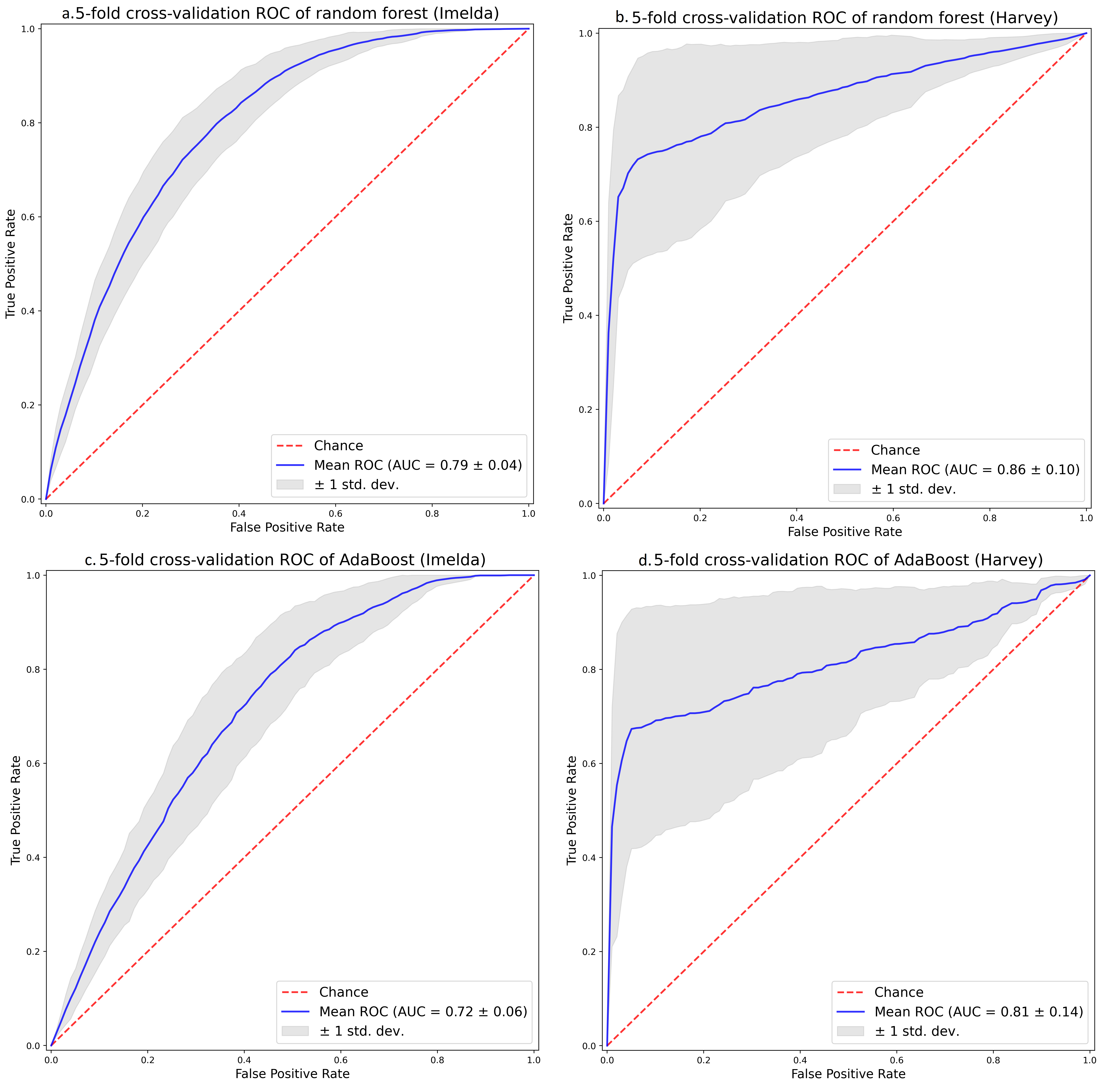}
\caption{Part 1: Receiver operating characteristic (ROC) curves for 10- iteration 5-fold cross-validation with random forest model for Tropical Storm Imelda (4a) and Hurricane Harvey (4b); Part 2: Receiver operating characteristic (ROC) curves for 10-fold cross-validation with AdaBoost model for Tropical Storm Imelda (4c) and Hurricane Harvey (4d).}
\end{figure}

In addition, both random forest and AdaBoost models performed better for predicting road inundation for Hurricane Harvey than for Tropical Storm Imelda (86.0 versus 81.0). This can be explained by the differences between Waze reports and IRINX traffic data. Waze reports are collected from Waze users, while IRINX traffic data was collected mainly by sensors. Humans are more sensitive to floods than are sensors, and as such, they may report flood risks on Waze when shallow water was found on the roads. Using roads with null values for average traffic speed to denote roads with flood risks in Hurricane Harvey is a more stringent standard (as most traffic data was collected by sensors) than the voluntary and subjective Waze reports. The advantage is that more roads with flood risks derived on Waze reports were still available for traffic use during Tropical Storm Imelda. 

Furthermore, our results demonstrate better performance than existing studies. Lee et al., (2017) used both random forest and boosted tree models to predict the spatial distribution of flood risks in the Seoul metropolitan area. They used the input features such as distance from the river (m), slope length factor (SLF), topographic wetness index (TWI), stream power index, and digital elevation model (DEM). Their regression computations of random forest and boosted tree models showed AUCs of 0.7878 and 0.7755, respectively. Compared with Lee et al., (2017), our random forest models had higher AUCs for both Hurricane Harvey (0.860) and Tropical Storm Imelda (0.790). For comparison of boosted tree and AdaBoost models, the AUC for Hurricane Harvey (0.810) is greater, while that of Tropical Storm Imelda (0.720) is less than that in Lee et al. (2017). The better performances of same (random forest) and similar (AdaBoost versus boosting tree) models in this study reflects the improved selection of input features such as the temporal precipitation features.

\subsection*{Model stability of random forest}
Given that the random forest model demonstrated better performance than the AdaBoost model, we further tested its stability with varying probability thresholds for detecting road inundation. The default probability threshold is 0.50: if the predicted flood probability of a road location is less than 0.50, the random forest model denotes this road as non-flooded class (i.e., negative class); otherwise, that road is categorized as flooded (i.e., positive class). Considering effective crisis response in floods, we are more concerned with the percentage of false negative predictions from our dataset. False negative refers to the result that roads with flood risks are incorrectly predicted as non-flooded roads. Larger false negative predictions could result in communities experiencing more severe flood impacts as residents may choose to evacuate via roads with false negative predictions (i.e., roads were predicted as non-flooded while they were actually flooded). Therefore, we tested the stability of the random forest model by adjusting the probability thresholds from 0.40 to 0.60 with a step of 0.01 and observing the variations of false negative percentages for both Tropical Storm Imelda and Hurricane Harvey. False negative percentage was computed using the Eq. (5). For each probability threshold, the 5-fold cross-validation process was implemented and the average false negative rate was calculated. With computed false negative rates, we plot their curves for Tropical Storm Imelda (green curve) and Hurricane Harvey (blue curve) in Figure 5.

\begin{equation}
\label{eq:5}
False \ negative \ percentage= \frac{False \ negative}{True \ positive +True \ negative +False \ positive+False \ negative}
\end{equation}

\begin{figure}[ht]
\centering
\includegraphics[width=0.6\linewidth]{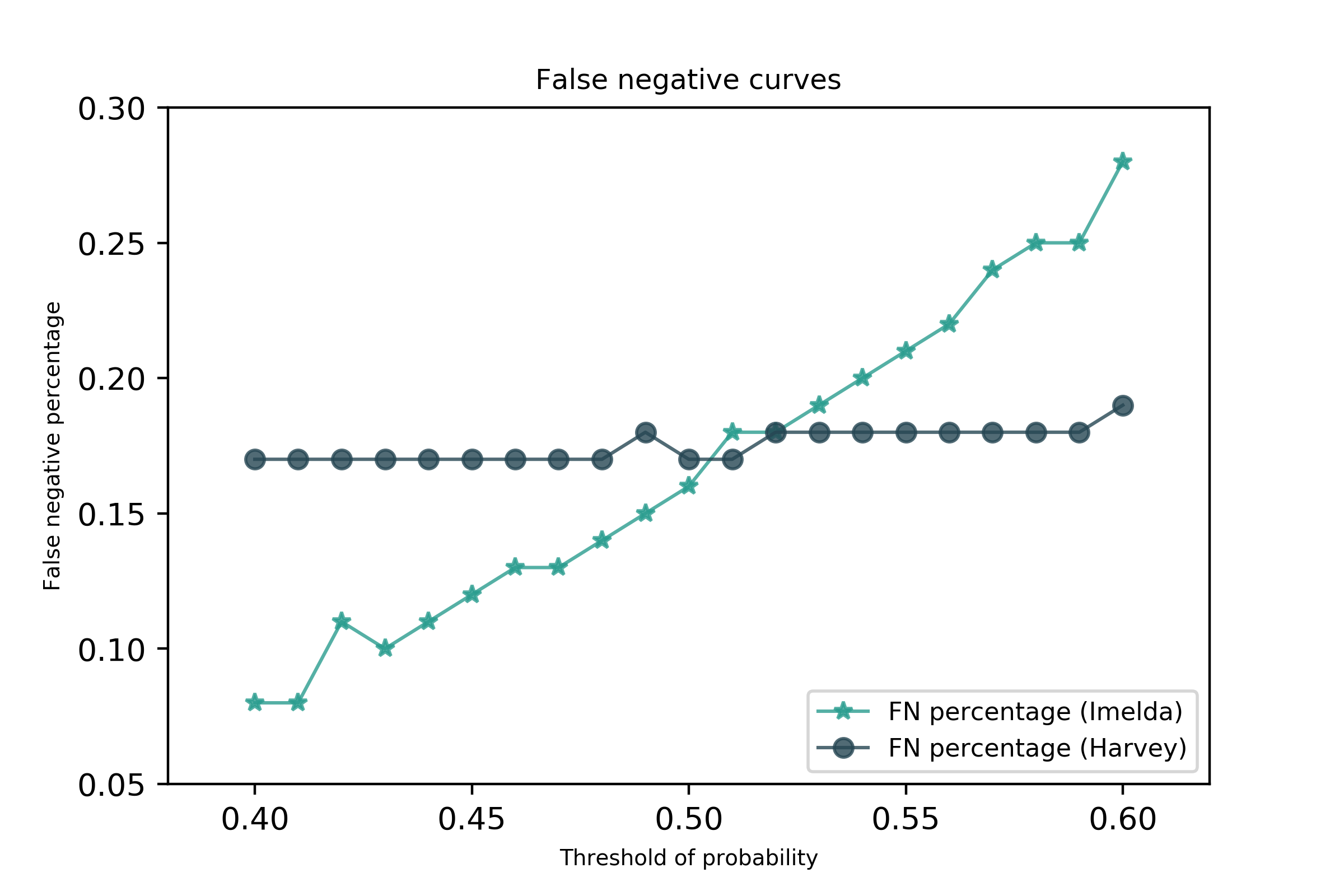}
\caption{False negative percentages from 5-fold cross-validation with random forest model by varying probability thresholds for Tropical Storm Imelda (green curve) and Hurricane Harvey (blue curve). FN means false negative.}
\end{figure}

Figure 5 reveals that the random forest model had stable performance for predicting road inundations during Hurricane Harvey. With probability threshold ranging from 0.40 to 0.60, we observed the false negative percentage changes from 0.17 to 0.19. For Tropical Storm Imelda, we discerned a significant variation of false negative percentage (from 0.08 to 0.28) when adjusting probability threshold from 0.40 to 0.60. When probability threshold is 0.50, the false negative percentage is 0.16 which is lower than the lower boundary of the false negative percentage range of Hurricane Harvey. Therefore, setting the probability threshold as 0.50 is reasonable for the random forest model for Hurricane Harvey, while selecting the probability threshold from 0.40 to 0.50 would yield a lower false negative percentage for Tropical Storm Imelda. In addition, precision and recall could also be a consideration when selecting the probability threshold (in terms of false negative percentage) for Tropical Storm Imelda.

\subsection*{Prediction results}
With 20\% of the test dataset used for both cases, we predicted the probabilities of these roads flooding using the random forest model (Figure 6). Nodes with varying colors from blue to red represent roads with low to high probabilities of getting flooded. Figure 6a (Tropical Storm Imelda) illustrates that roads at high risk of being flooded are mainly in the center of Houston (Imelda), while Figure 6b (Hurricane Harvey) shows those with high flood-risk levels mainly surround the boundary of Harris County. Meyerland (green-shaded region in Figure 6b) is almost entirely located within the 100-year floodplain and was inundated in Hurricane Harvey. From the predicted probability in Figure 6b, we can see one road within in this region has the probability of 57.30\% (medium level) to be inundated during Hurricane Harvey. 

\begin{figure}[ht]
\centering
\includegraphics[width=1.0\linewidth]{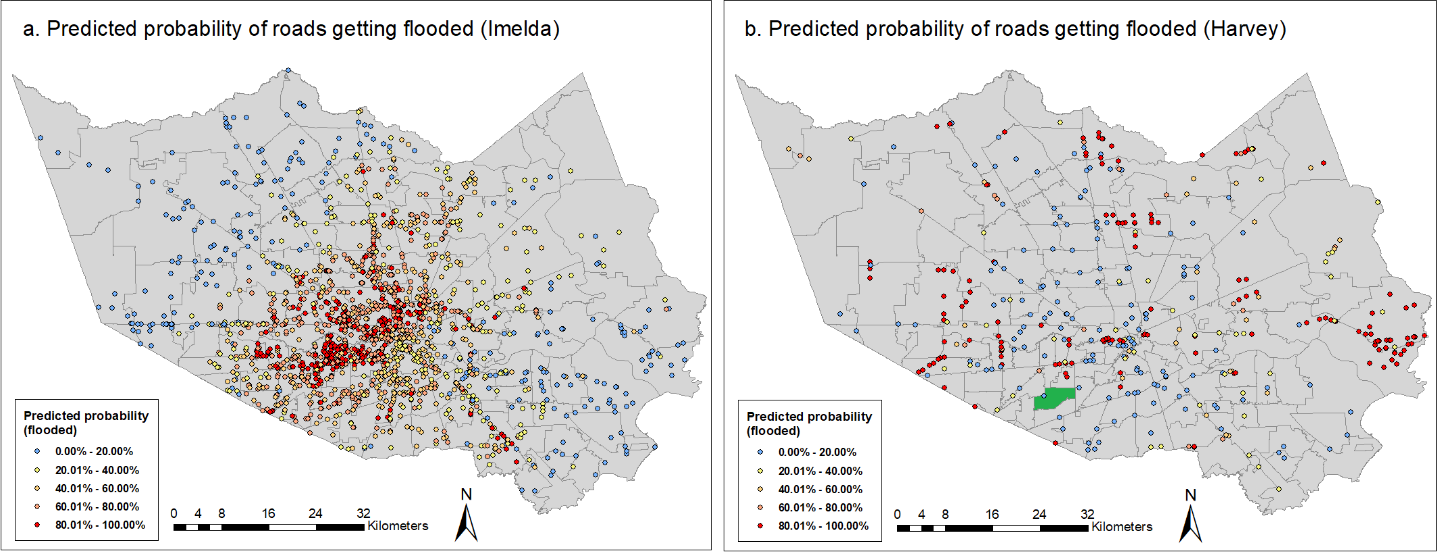}
\caption{Examples of predictions of probabilities for roads becoming flooded in Tropical Storm Imelda (6a) and Hurricane Harvey (6b). The green-shaded polygon represents the neighborhood of Meyerland in southwest Houston. Probabilities denote the risk levels of roads getting flooded: 0.00\%–20.00\%: very low; 20.01\%–40.00\%: low; 40.01\%–60.00\%: medium; 60.01\%–80.00\%: high; and 80.01\%–100.00\%: very high.}
\end{figure}

With the predicted probabilities, we can categorize roads into varying risk levels. For instance, roads shaded red can be categorized as being at very high risk of flooding, while those shaded blue are at very low flood risk. Residents of impacted communities can choose their vehicle trips or evacuation routes by avoiding roads with very high flood-risk levels. Emergency management agencies can identify communities losing access to essential facilities (such as hospitals) and prioritize delivery of relief resources to these communities. In addition, our model makes possible a tool which can help design scenarios by considering varying probabilities for roads getting flooded. Compared with the flood scenarios of 100-year and 500-year floodplains, this tool integrates not only the latest topographic and hydrologic features, but also temporal precipitation features. Therefore, the design of flood scenarios with our model in this study could have better representation of the actual conditions.

According to the predicted probabilities (Figures 6a and 6b), we can denote the predictive flood status of roads with proper probability thresholds. Referring to Figure 5, we used probability thresholds of 0.45 and 0.50 to detect inundated roads (i.e., positive class) during Tropical Storm Imelda and Hurricane Harvey, respectively. Then, we showed examples of the prediction results for both cases in Figure 7, where the green links represent roads with true positive predictions, red for false negative predictions, brown for false positive predictions, and blue for true negative predictions.

\begin{figure}[ht]
\centering
\includegraphics[width=1.0\linewidth]{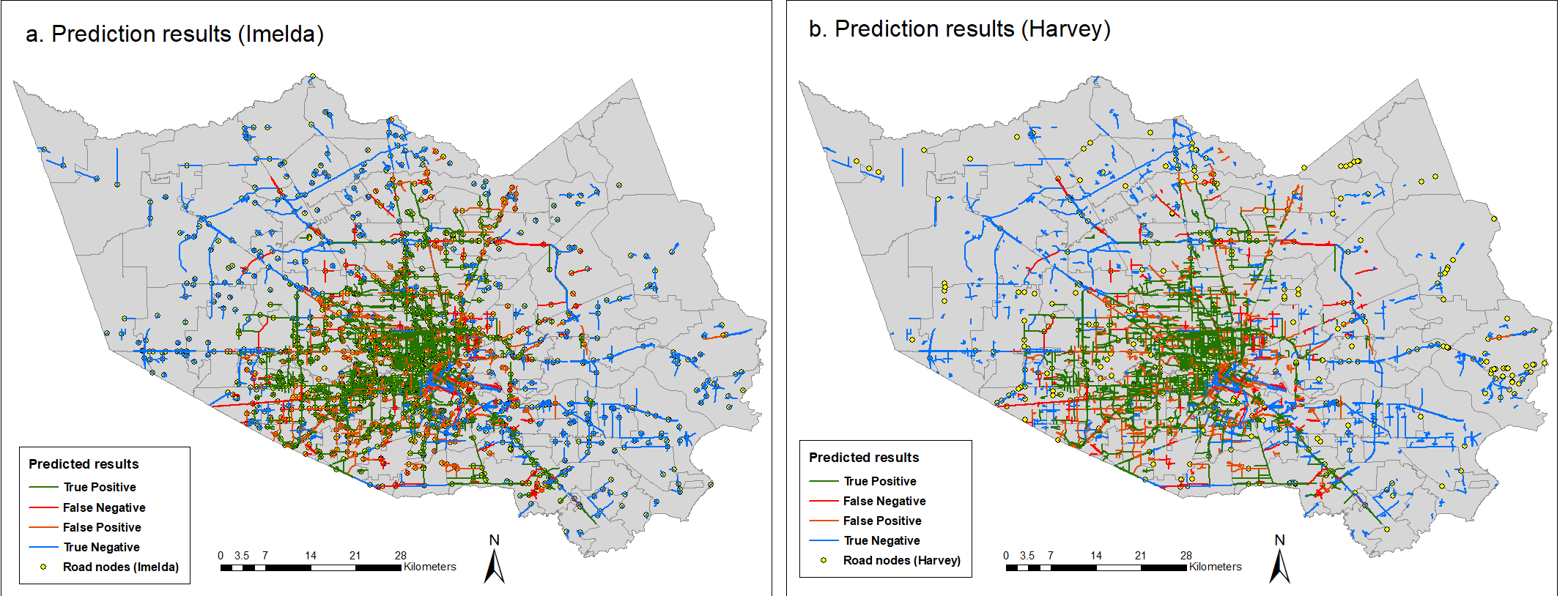}
\caption{Examples of prediction results for roads with and without flood risks for Tropical Storm Imelda (7a) and Hurricane Harvey (7b) according to the predicted probabilities in Figures 6a and 6b, respectively. Each link represents the road with road node used for model test. For better presentations, we show only the roads used in our test dataset.}
\end{figure}

\section*{Discussions and Concluding Remarks}

The results demonstrate that roads with high flood-risk potential can be accurately predicted using machine learning models in the context of two storm cases in Harris County. For predicting road flooding, precipitation features for extreme storm events (Hurricane Harvey) are more important predictors than topographic and hydrologic features for predicting flooded roads. In addition to precipitation, topographic features (elevation, coastal and stream proximity) and height above nearest drainage generally have greater influence than hydrologic features (roughness, imperviousness, and saturated hydraulic conductivity) for causing road inundations, which is generally consistent with the results reported by Mobley et al., (2021) and Lee et al., (2017). Compared with existing studies using random forest and boosted tree models for flood risk predictions (Lee et al., 2017), our corresponding models demonstrate higher AUC values. This difference is likely to be explained by the addition of temporal precipitation features (total rainfall in 1 hour, 2 hours, and 24 hours) as inputs for predicting flooded roads in our study, particularly for storm events with extreme precipitation such as Hurricane Harvey. Overall, our two models present strong predictive capability for detecting roads at risk for flooding. The random forest model (bagging technique-based algorithm) demonstrates better performance than AdaBoost mode (boosting technique-based algorithm). 

As we are concerned with reducing false negative predictions (flooded roads incorrectly predicted as non-flooded), we use the false negative percentage with varying probability thresholds to test model stability. The results show that the random forest model has stable performances for Hurricane Harvey with minor variations of false negative percentage when adjusting probability thresholds but show significant variations across false negative percentage for Tropical Storm Imelda. This difference could be attributed to the dependent variables used for identifying flooded roads. The dataset for Tropical Storm Imelda was drawn from Waze reports, which are compiled from crowdsourced data. These users were more sensitive to road flooding than are traffic sensors; for instance, roads with shallow water pooling could be still available for vehicle use, but Waze users may report flood warnings for these roads. Roads which may not be severely inundated (still passable) during Tropical Storm Imelda were more likely to be labeled as flooded roads, which can further impact the stability of model performances. IRINX traffic data from which Hurricane Harvey dataset was drawn, were mostly collected by sensors. We proceeded on the assumption that roads with null values for average traffic speed during Hurricane Harvey were unused, and therefore flooded. This study was constrained by the use of two dissimilar datasets which prevents direct comparison between road conditions evident during Hurricane Harvey and Tropical Storm Imelda. 

The study and findings contribute to the emerging field of smart flood resilience focusing on harnessing community-scale big data and machine learning approaches to enhance disaster resilience capabilities, such as predictive flood risk mapping at the road level (Dong et al., 2020d). In addition, this study provides a tool to detect roads with varying flood risk levels using topographic, hydrologic, and precipitation features (Figure 6). This tool can be integrated with a percolation analysis of the road network (Dong et al., 2021, 2020d) so that the removal of roads can refer to the roads with higher predicted probability of getting flooded (Li et al., 2015), which is a more precise method than relying on 100-year and 500-year floodplain maps. For instance, roads with higher flood risk can be removed from the road network (for network analysis) to explore its vulnerability. Considering the variations of feature importance for Tropical Storm Imelda and Hurricane Harvey (Figure 3), for instance Hurricane Harvey brought more extreme rainfalls, our results suggest using the random forest model (trained on Tropical Storm Imelda) for flood events without abrupt severe rainfalls and using the model trained on Hurricane Harvey for floods with slow-moving extreme rainfalls. 

Our findings can help potentially impacted communities identify roads that are more likely to be inundated by hurricanes and floods. This foresight could be incorporated in navigation applications to help drivers avoid roads with high flooding probability when accessing essential facilities such as hospitals and groceries. Incidents from the past events indicate that driving through flooded roads is among the leading cause of deaths during urban floods (Jonkman and Kelman 2005; Fitzgerald et al., 2010; Drobot et al., 2007); our model could help affected residents avoid driving into flooded roads. This study also provides a tool for emergency management agencies (EMAs) to design response strategies, such as evacuation plans  before the arrival of hurricanes and floods. For instance, with the predicted probability map for road flooding risk (Figure 6), EMAs could identify which community is more likely to lose accessibility to critical facilities such as medical facilities and groceries. For residents in those communities, EMAs could notify and help them evacuate to shelters and other safe places in the short-term preparedness stage before hurricanes and floods approach. For roads connecting communities to critical facilities and having high probability of being inundated, EMAs can inform the road infrastructure operators and maintainers (such as Texas Department of Transportation) to take protection actions, such as sandbags, along these roads. Furthermore, our model could help EMAs update floodplain maps with more specific infrastructure risk information, which could guide the urban plan strategies for future flood hazard mitigation; with for instance, with tropical and hydrologic features, and historical precipitation data in a city, our model could help predict the probability of different road sections to be flooded and then produce the floodplain map with infrastructure risk information for that region. Compared with traditional 100- and 500-year floodplain maps that are missing infrastructure risk insights, those from our models can be more easily produced and updated with updated topographic, hydraulic and precipitation features. As a result, infrastructure risk insights from our models could better reflect the updated topographic and hydraulic features and the changes to infrastructure risks could be incorporated into 100- and 500-year floodplain maps. 

The study has some limitations. For both models, sample selections of non-flooded roads could impact their performance. A proper selection method for identifying non-flooded roads plays a critical role for improving model performances (Darabi et al., 2019; Barbet-Massin et al., 2012). In this study, we used Waze reports to identify roads with flood risks in Tropical Storm Imelda, but there is no comparable record for roads without flood risks. Using roads from the IRINX traffic dataset (and excluded roads with flood warnings from Waze reports), we randomly selected equivalent-size of roads without flood warnings. For comparison considerations (both storm cases have balanced datasets), we also randomly selected the equivalent-size of roads with traffic speed value for Hurricane Harvey. Future work will focus on the investigation of the impacts from random selection of non-flooded roads on model performance. In addition, we used two different datasets for two different storm cases due to limited data availability. Waze reports data became available after 2018, so Waze data was not available for 2017 Hurricane Harvey. IRINIX traffic data for Tropical Storm Imelda was not available to the research team. Using different datasets could impact the model performance due to their different natures as mentioned in earlier in this section; however, our results demonstrate that different datasets could be used for training machine learning models for predicting road flooding risk, and future users can choose their datasets based on availability.

Another limitation to be noted is the assignment of topographic and hydrologic features to roads. Features such as roughness and imperviousness attributed to the roads for both Tropical Storm Imelda (2019) and Hurricane Harvey (2017) came from the closest year for which the data were made available. In our models, both storm cases used roughness and imperviousness from the 2016 National Land Cover Database, although these variables were heavily impacted by large-scale changes in urbanization over time. The consequence is that the roughness and imperviousness are less representative of actual hydrologic conditions in 2017 and 2019, which may explain the low-level rank of their importance for predicting roads with flood risks (Figure 3). Future work upon availability of updated data could consider urban development by using proper methods to adjust the features which are not available in the disaster years.

Despite limitations, this study provides a model to predict road flooding risk with topographic, hydrologic, and precipitation features. With crowdsourced (Waze reports) and sensor (IRINX traffic) data, the model is trained and tested with high accuracy for predicting roads with flood risk. This modeling cannot only be generalized to other flood events and regions with proper topographic, hydrologic, and precipitation features for predicting flooded roads, but also be used as a tool to design road failure scenarios (roads with predictive probability of getting flooded) for percolation analysis of road network. Our model can also benefit potentially impacted communities and emergency management agencies’ preparedness and response actions to hurricanes and floods. Accordingly, the model and results contribute to the emerging field of smart flood resilience (Fan et al., 2021) aiming to harness heterogeneous datasets to improve situational awareness and predictive monitoring during disasters. 


\section*{Acknowledgements}
The authors would like to acknowledge the funding support from the X-Grant program (Presidential Excellence Fund) from the Texas A\&M University. The authors would also like to acknowledge INRIX, and WAZE for providing the traffic data and WAZE report data respectively. Any opinions, findings, and conclusions or recommendations expressed in this research are those of the authors and do not necessarily reflect the views of the funding agencies.

\section*{Data availability}
The data that support the findings of this study are available from Waze and INRIX, but restrictions apply to the availability of these data, which were used under license for the current study. The data can be accessed upon request submitted on each data provider. Other data (flood inundations and flood claims) we use in this study are all publicly available.

\section*{Code availability}
The code that supports the findings of this study is available from the corresponding author upon request.

\section*{References}
Acrement, G. J., \& Schneider, V. R. (1984). Guide for Selecting Manning’s Roughness Coefficients for Natural Channels and flood plains. Federal Highways Administration, US Department of Transportation (p. 66). Report No. FHWA-TS-84-204, Washington.

Anderson, B. G., Rutherfurd, I. D., \& Western, A. W. (2006). An analysis of the influence of riparian vegetation on the propagation of flood waves. Environmental Modelling \& Software, 21(9), 1290-1296.

Anderson, D. G. (1970). Effects of urban development on floods in northern Virginia (p. 22). US Government Printing Office.

Arnold Jr, C. L., \& Gibbons, C. J. (1996). Impervious surface coverage: the emergence of a key environmental indicator. Journal of the American planning Association, 62(2), 243-258.

Barbet‐Massin, M., Jiguet, F., Albert, C. H., \& Thuiller, W. (2012). Selecting pseudo‐absences for species distribution models: how, where and how many?. Methods in ecology and evolution, 3(2), 327-338.

Bates, P. D., \& De Roo, A. P. J. (2000). A simple raster-based model for flood inundation simulation. Journal of hydrology, 236(1-2), 54-77.

Blake, E., \& Zelinsky, D. (2018). NATIONAL HURRICANE CENTER TROPICAL CYCLONE REPORT: HURRICANE HARVEY. Available at <https://www.nhc.noaa.gov/data/tcr/AL092017\_Harvey.pdf>, last accessed on June 8, 2021.

Bui, D. T., Ngo, P. T. T., Pham, T. D., Jaafari, A., Minh, N. Q., Hoa, P. V., \& Samui, P. (2019). A novel hybrid approach based on a swarm intelligence optimized extreme learning machine for flash flood susceptibility mapping. Catena, 179, 184-196.

Brody, S. D., Highfield, W. E., \& Blessing, R. (2015). An analysis of the effects of land use and land cover on flood losses along the Gulf of Mexico coast from 1999 to 2009. JAWRA Journal of the American Water Resources Association, 51(6), 1556-1567.

Chaudhry, M. A., Naeem, U. A., \& Hashmi, H. N. (2018). Performance evaluation of 1-D numerical model HEC-RAS towards modeling sediment depositions and sediment flushing operations for the reservoirs. Environmental monitoring and assessment, 190(7), 1-18.

Chen, A. S., Djordjevic, S., Leandro, J., \& Savic, D. (2007). The urban inundation model with bidirectional flow interaction between 2D overland surface and 1D sewer networks. In Novatech 2007-6ème Conférence sur les techniques et stratégies durables pour la gestion des eaux urbaines par temps de pluie/Sixth International Conference on Sustainable Techniques and Strategies in Urban Water Management. GRAIE, Lyon, France.

Chen, W., Huang, G., Zhang, H., \& Wang, W. (2018). Urban inundation response to rainstorm patterns with a coupled hydrodynamic model: A case study in Haidian Island, China. Journal of Hydrology, 564, 1022-1035.

Coles, D., Yu, D., Wilby, R. L., Green, D., \& Herring, Z. (2017). Beyond ‘flood hotspots’: Modelling emergency service accessibility during flooding in York, UK. Journal of Hydrology, 546, 419-436.

Coltin, B., McMichael, S., Smith, T., \& Fong, T. (2016). Automatic boosted flood mapping from satellite data. International Journal of Remote Sensing, 37(5), 993-1015.

Cook, A., \& Merwade, V. (2009). Effect of topographic data, geometric configuration and modeling approach on flood inundation mapping. Journal of Hydrology, 377(1-2), 131-142.

Darabi, H., Choubin, B., Rahmati, O., Haghighi, A. T., Pradhan, B., \& Kløve, B. (2019). Urban flood risk mapping using the GARP and QUEST models: A comparative study of machine learning techniques. Journal of hydrology, 569, 142-154.

Dodangeh, E., Choubin, B., Eigdir, A. N., Nabipour, N., Panahi, M., Shamshirband, S., \& Mosavi, A. (2020). Integrated machine learning methods with resampling algorithms for flood susceptibility prediction. Science of the Total Environment, 705, 135983.

Dong, S., Malecha, M., Farahmand, H., Mostafavi, A., Berke, P. R., \& Woodruff, S. C. (2021). Integrated infrastructure-plan analysis for resilience enhancement of post-hazards access to critical facilities. Cities, 117, 103318.

Dong, S., Esmalian, A., Farahmand, H., \& Mostafavi, A. (2020a). An integrated physical-social analysis of disrupted access to critical facilities and community service-loss tolerance in urban flooding. Computers, Environment and Urban Systems, 80, 101443.

Dong, S., Li, Q., Farahmand, H., Mostafavi, A., Berke, P. R., \& Vedlitz, A. (2020b). Institutional connectedness in resilience planning and management of interdependent infrastructure systems. Journal of Management in Engineering, 36(6), 04020075.

Dong, S., Yu, T., Farahmand, H., \& Mostafavi, A. (2020c). Bayesian modeling of flood control networks for failure cascade characterization and vulnerability assessment. Computer‐Aided Civil and Infrastructure Engineering, 35(7), 668-684.

Dong, S., Yu, T., Farahmand, H., \& Mostafavi, A. (2020d). Probabilistic modeling of cascading failure risk in interdependent channel and road networks in urban flooding. Sustainable Cities and Society, 62, 102398.

Drobot, S. D., Benight, C., \& Gruntfest, E. C. (2007). Risk factors for driving into flooded roads. Environmental Hazards, 7(3), 227-234.

Fan, C., Zhang, C., Yahja, A., \& Mostafavi, A. (2021). Disaster City Digital Twin: A vision for integrating artificial and human intelligence for disaster management. International Journal of Information Management, 56, 102049.

Fan, C., Jiang, X., \& Mostafavi, A. (2020). A network percolation-based contagion model of flood propagation and recession in urban road networks. Scientific Reports, 10(1), 1-12.

FitzGerald, G., Du, W., Jamal, A., Clark, M., \& Hou, X. Y. (2010). Flood fatalities in contemporary Australia (1997–2008). Emergency Medicine Australasia, 22(2), 180-186.

Garousi‐Nejad, I., Tarboton, D. G., Aboutalebi, M., \& Torres‐Rua, A. F. (2019). Terrain analysis enhancements to the height above nearest drainage flood inundation mapping method. Water Resources Research, 55(10), 7983-8009.

Gori, A., Blessing, R., Juan, A., Brody, S., \& Bedient, P. (2019). Characterizing urbanization impacts on floodplain through integrated land use, hydrologic, and hydraulic modeling. Journal of hydrology, 568, 82-95.
Hall, M. J. (1984). Urban hydrology. Elsevier Applied Science Publishing. London, (37474), 299.

Hosmer, D. W., \& Lemeshow, S. (2000). Logistic Regression 2nd edn John Wiley and Sons. Inc.: New York.

Hosseini, F. S., Choubin, B., Mosavi, A., Nabipour, N., Shamshirband, S., Darabi, H., \& Haghighi, A. T. (2020). Flash-flood hazard assessment using ensembles and Bayesian-based machine learning models: Application of the simulated annealing feature selection method. Science of the total environment, 711, 135161.

Hou, J., Zhou, N., Chen, G. et al., Rapid forecasting of urban flood inundation using multiple machine learning models. Nat Hazards (2021). https://doi.org/10.1007/s11069-021-04782-x

Huang, Q., Wang, J., Li, M., Fei, M., \& Dong, J. (2017). Modeling the influence of urbanization on urban pluvial flooding: a scenario-based case study in Shanghai, China. Natural Hazards, 87(2), 1035-1055.

Jamali, B., Bach, P. M., Cunningham, L., \& Deletic, A. (2019). A Cellular Automata fast flood evaluation (CA‐ffé) model. Water Resources Research, 55(6), 4936-4953.

Jamali, B., Löwe, R., Bach, P. M., Urich, C., Arnbjerg-Nielsen, K., \& Deletic, A. (2018). A rapid urban flood inundation and damage assessment model. Journal of Hydrology, 564, 1085-1098.

Janizadeh, S., Avand, M., Jaafari, A., Phong, T. V., Bayat, M., Ahmadisharaf, E., ... \& Lee, S. (2019). Prediction success of machine learning methods for flash flood susceptibility mapping in the tafresh watershed, Iran. Sustainability, 11(19), 5426.

Jonkman, S. N., \& Kelman, I. (2005). An analysis of the causes and circumstances of flood disaster deaths. Disasters, 29(1), 75-97.

Kalyanapu, A. J., Burian, S. J., \& McPherson, T. N. (2009). Effect of land use-based surface roughness on hydrologic model output. Journal of Spatial Hydrology, 9(2).

Karaoui, I., Arioua, A., El Amrani Idrissi, A., Hssaisoune, M., Nouaim, W., Ait ouhamchich, K., \& Elhamdouni, D. (2018). Assessing land use/cover variation effects on flood intensity via hydraulic simulations: A case study of Oued El Abid watershed (Morocco). Geology, Ecology, and Landscapes, 2(2), 73-80.

Latto, A., \& Berg, R. (2020). NATIONAL HURRICANE CENTER TROPICAL CYCLONE REPORT: TROPICAL STORM IMELDA. Available at < https://www.nhc.noaa.gov/data/tcr/AL112019\_Imelda.pdf>, last accessed on June 8, 2021.

Lee, C. C. B., \& Gharaibeh, N. G. (2020). Automating the evaluation of urban roadside drainage systems using mobile lidar data. Computers, Environment and Urban Systems, 82, 101502.

Lee, S., Kim, J. C., Jung, H. S., Lee, M. J., \& Lee, S. (2017). Spatial prediction of flood susceptibility using random-forest and boosted-tree models in Seoul metropolitan city, Korea. Geomatics, Natural Hazards and Risk, 8(2), 1185-1203.

Li, Q., Yang, Y., Wang, W., Lee, S., Xiao, X., Gao, X., ... \& Mostafavi, A. (2021). Unraveling the Dynamic Importance of County-level Features in Trajectory of COVID-19. arXiv preprint arXiv:2101.03458.

Li, D., Zhang, Q., Zio, E., Havlin, S., \& Kang, R. (2015). Network reliability analysis based on percolation theory. Reliability Engineering \& System Safety, 142, 556-562.

Liaw, A., \& Wiener, M. (2002). Classification and regression by randomForest. R news, 2(3), 18-22.

Liu, L., Liu, Y., Wang, X., Yu, D., Liu, K., Huang, H., \& Hu, G. (2015). Developing an effective 2-D urban flood inundation model for city emergency management based on cellular automata. Natural hazards and earth system sciences, 15(3), 381-391.

Liu, X., Sahli, H., Meng, Y., Huang, Q., \& Lin, L. (2017). Flood inundation mapping from optical satellite images using spatiotemporal context learning and modest AdaBoost. Remote Sensing, 9(6), 617.

Liu, Y. Y., Maidment, D. R., Tarboton, D. G., Zheng, X., Yildirim, A., Sazib, N. S., \& Wang, S. (2016). A CyberGIS approach to generating high-resolution height above nearest drainage (HAND) raster for national flood mapping.

Lyu, H. M., Shen, S. L., Yang, J., \& Yin, Z. Y. (2019). Inundation analysis of metro systems with the storm water management model incorporated into a geographical information system: a case study in Shanghai. Hydrology and Earth System Sciences, 23(10), 4293-4307.

Mobley, W., Sebastian, A., Blessing, R., Highfield, W. E., Stearns, L., \& Brody, S. D. (2021). Quantification of continuous flood hazard using random forest classification and flood insurance claims at large spatial scales: a pilot study in southeast Texas. Natural Hazards and Earth System Sciences, 21(2), 807-822.

Mobley, W., Sebastian, A., Highfield, W., \& Brody, S. D. (2019). Estimating flood extent during Hurricane Harvey using maximum entropy to build a hazard distribution model. Journal of Flood Risk Management, 12, e12549.

Nguyen, D. H., \& Bae, D. H. (2020). Correcting mean areal precipitation forecasts to improve urban flooding predictions by using long short-term memory network. Journal of Hydrology, 584, 124710.

NOAA (National Oceanic and Atmospheric Administration). (2017). Fast facts: Hurricane costs. Silver Spring, MD: NOAA.

Nobre, A. D., Cuartas, L. A., Hodnett, M., Rennó, C. D., Rodrigues, G., Silveira, A., \& Saleska, S. (2011). Height Above the Nearest Drainage–a hydrologically relevant new terrain model. Journal of Hydrology, 404(1-2), 13-29.

Oza, N. C., \& Russell, S. J. (2001). Online bagging and boosting. In International Workshop on Artificial Intelligence and Statistics (pp. 229-236). PMLR.

Podesta, C., Coleman, N., Esmalian, A., Yuan, F., \& Mostafavi, A. (2021). Quantifying community resilience based on fluctuations in visits to points-of-interest derived from digital trace data. Journal of the Royal Society Interface, 18(177), 20210158.

Prasad, A. M., Iverson, L. R., \& Liaw, A. (2006). Newer classification and regression tree techniques: bagging and random forests for ecological prediction. Ecosystems, 9(2), 181-199.

Qian, Z. (2010). Without zoning: Urban development and land use controls in Houston. Cities, 27(1), 31-41.

Rahmati, O., \& Pourghasemi, H. R. (2017). Identification of critical flood prone areas in data-scarce and ungauged regions: a comparison of three data mining models. Water resources management, 31(5), 1473-1487.

Rodda, H. J. (2005). The development and application of a flood risk model for the Czech Republic. Natural hazards, 36(1), 207-220.

Rawls, W. J., Brakensiek, D. L., \& Miller, N. (1983). Green-Ampt infiltration parameters from soils data. Journal of hydraulic engineering, 109(1), 62-70.

Schapire, R. E. (2013). Explaining adaboost. In Empirical inference (pp. 37-52). Springer, Berlin, Heidelberg.

Sebastian, A., Gori, A., Blessing, R. B., van der Wiel, K., \& Bass, B. (2019). Disentangling the impacts of human and environmental change on catchment response during Hurricane Harvey. Environmental Research Letters, 14(12), 124023.

Smith, R. A., Bates, P. D., \& Hayes, C. (2012). Evaluation of a coastal flood inundation model using hard and soft data. Environmental Modelling \& Software, 30, 35-46.

Sutton, C. D. (2005). Classification and regression trees, bagging, and boosting. Handbook of statistics, 24, 303-329.

Tehrany, M. S., Kumar, L., \& Shabani, F. (2019). A novel GIS-based ensemble technique for flood susceptibility mapping using evidential belief function and support vector machine: Brisbane, Australia. PeerJ, 7, e7653.

Thomas, H., \& Nisbet, T. R. (2007). An assessment of the impact of floodplain woodland on flood flows. Water and Environment Journal, 21(2), 114-126.

Versini, P. A. (2012). Use of radar rainfall estimates and forecasts to prevent flash flood in real time by using a road inundation warning system. Journal of hydrology, 416, 157-170.

White, M. D., \& Greer, K. A. (2006). The effects of watershed urbanization on the stream hydrology and riparian vegetation of Los Penasquitos Creek, California. Landscape and urban Planning, 74(2), 125-138.

Yin, J., Yu, D., Yin, Z., Liu, M., \& He, Q. (2016a). Evaluating the impact and risk of pluvial flash flood on intra-urban road network: A case study in the city center of Shanghai, China. Journal of hydrology, 537, 138-145.

Yin, J., Yu, D., \& Wilby, R. (2016). Modelling the impact of land subsidence on urban pluvial flooding: A case study of downtown Shanghai, China. Science of the Total Environment, 544, 744-753.

Yu, D., Yin, J., \& Liu, M. (2016). Validating city-scale surface water flood modelling using crowd-sourced data. Environmental Research Letters, 11(12), 124011.

Yuan, F., Liu, R., Mao, L., \& Li, M. (2021a). Internet of people enabled framework for evaluating performance loss and resilience of urban critical infrastructures. Safety Science, 134, 105079.

Yuan, F., Xu, Y., Li, Q., \& Mostafavi, A. (2021b). Spatio-Temporal Graph Convolutional Networks for Road Network Inundation Status Prediction during Urban Flooding. arXiv preprint arXiv:2104.02276.

Yuan, F., Yang, Y., Li, Q., \& Mostafavi, A. (2021c). Unraveling the Temporal Importance of Community-scale Human Activity Features for Rapid Assessment of Flood Impacts. arXiv preprint arXiv:2106.08370.

\end{document}